\documentclass[10pt,journal]{IEEEtran}
\usepackage{multirow}
\usepackage{graphicx}
\usepackage{mathtools}
\usepackage{commath}
\usepackage{amsmath}
\usepackage{graphicx}
\usepackage{algorithm}
\usepackage{algpseudocode}
\usepackage{algorithmicx}
\usepackage{epstopdf}
\usepackage{multirow}
\usepackage{array}
\usepackage{pgfplots}

\DeclareMathAlphabet\mathbfcal{OMS}{cmsy}{b}{n} 
\usepackage{amssymb} 
\usepackage{mathtools}

\usepackage{extarrows}

\DeclarePairedDelimiter\floor{\lfloor}{\rfloor}
\begin{document}
\title{Tensor-Train Long Short-Term Memory for Monaural Speech Enhancement}

\author{Suman~Samui,
        Indrajit~Chakrabarti,
        and~Soumya~K.~Ghosh,
\thanks{S. Samui is with Idiap Research Institute, Martigny, Switzerland and also with Indian Institute of Technology (IIT) Kharagpur, India (e-mail: ssamui@idiap.ch), I. Chakrabarti is with Department of Electronics and Electrical Communication Engineering, IIT Kharagpur (e-mail: indrajit@ece.iitkgp.ac.in), and S. K. Ghosh is with Department of Computer Science and Engineering, IIT Kharagpur, West Bengal-721302, India (e-mail: skg@iitkgp.ac.in).}}

\maketitle

\begin{abstract}
In recent years, Long Short-Term Memory (LSTM) has become a popular choice for speech separation and speech enhancement task. The capability of LSTM network can be enhanced by widening and adding more layers. However, this would introduce millions of parameters in the network and also increase the requirement of computational resources. These limitations hinders the efficient implementation of RNN models in low-end devices such as mobile phones and embedded systems with limited memory. To overcome these issues, we proposed to use an efficient alternative approach of reducing parameters by representing the weight matrix parameters of LSTM based on Tensor-Train (TT) format. We called this Tensor-Train factorized LSTM as TT-LSTM model. Based on this TT-LSTM units, we proposed a deep TensorNet model for single-channel speech enhancement task. Experimental results in various test conditions and in terms of standard speech quality and intelligibility metrics, demonstrated that the proposed deep TT-LSTM based speech enhancement framework can achieve competitive performances with the state-of-the-art uncompressed RNN model, even though the proposed model architecture is orders of magnitude less complex.

\end{abstract}

\begin{IEEEkeywords}
Speech enhancement, LSTM, Tensor-Train.
\end{IEEEkeywords}

\section{Introduction}
Supervised data-driven approaches of speech enhancement have gained  a significant research attention in recent years. Unlike traditional methods based on statistical signal processing \cite{loizou2013}, these machine learning based methods could able to deal with the non-stationary noise conditions and perform comparatively well in transient noise scenarios \cite{sie2017}. 
In this work, our main focus is on the single-channel speech enhancement task, i.e., it would be assumed that the noisy speech samples captured by single-channel microphone are available at the input of system.

In the last few years, a plethora of methods with different types of machine learning architectures have been investigated towards achieving better generalization in various dimensions such as Noise-type, speaker and SNR level, mainly involving with speech enhancement task \cite{sie2017}\cite{chen2017long}.  Long Short Term Memory (LSTM) unit \cite{hochreiter1997long}\cite{greff2017lstm} based  Recurrent Neural Network (RNN) is  a prominent choice in solving speech enhancement just like the other various natural language processing tasks such as speech recognition \cite{graves2013speech} and machine translation \cite{kalchbrenner2013recurrent}. In \cite{chen2017long}, it has been reported that LSTM can able to achieve better speaker generalization than the fully connected deep neural network. The popularity of LSTM is mainly because of its ease of optimization than vanilla-RNN). LSTM uses stochastic gradient descent, but changes the hidden units in such a way that the backpropagated gradients are much better behaved \cite{lipton2015critical}. The initial proposed structure of LSTM \cite{hochreiter1997long} replaces logistic or tanh hidden units with ``memory cells" which has  its own input and output gates that control when inputs are allowed to add to the stored analog value and when this value is allowed to influence the output.  The subsequent introduction of forget gates in LSTM help to learn long-range dependencies in a more better way. Although this complex gated RNN architectures can be easily optimized by gradient based optimization techniques, we need many dense matrices to represent a basic LSTM unit. In fact, a LSTM layer has more than four times parameters than a vanila-RNN layer. Moreover, stacking of several LSTM layers introduce millions of more parameters in the network and also increase the requirement of computational resources, making it infeasible to deploy such model in the low-end devices such as mobile phones and embedded systems with limited memory. To address these issues, we introduced Tensor-Train LSTM (TT-LSTM) where the weight matrices of LSTM are decomposed in Tensor-Train format (TTF). This type of TTF representation are able to provide significant compression in LSTM representation. We proposed a deep $TensorNet$ with three hidden layers of TT-LSTM for speech enhancement task.  The proposed framework adopted a T-F masking based speech enhancement approach just as in \cite{chen2017long}\cite{samui2017deep}\cite{samui2018asoc}. Extensive experiments in various noise conditions have shown that TT-LSTM based deep TensorNet are able to provide competitive performances with the state-of-the-art uncompressed RNN model, even though the proposed model architecture is orders of magnitude less complex.

The remainder of the paper is organized as follows. In Section 2, we introduce Tensor-Train format representation and also derived TT-LSTM model. Section 3 presents the proposed speech enhancement framework. Experimental results are described in Section 4. Finally, Section 5 concludes the work. 

\section{Tensor-Train format representation}
Tensor-train format was first introduced in \cite{oseledets2011tensor} where the authors presented a simple non-recursive form of the tensor decomposition method with the advantage of being capable of scaling to an arbitrary number of dimensions. Recently, it has been also shown  that a deep neural network can be effectively compressed by converting its dense weight matrix of fully connected layer into a high-dimensional tensor-train format \cite{novikov2015tensorizing}.  

In the following subsections, we first introduce the tensor-train factorization method and then show how this same method can be applied to represent the weight matrices of a gated RNN model such as LSTM.   
\subsection{Tensor-Train factorization (TTF)}
Tensor-train factorization model is flexible in the sense that it can scale to an arbitrary number of dimensions. A $d$-dimensional tensor of the form $\mathbfcal{A}$ $\in$ $\mathbb{R}^{t_1 \times t_2\dots\times t_{(d-1)}\times t_d }$ can be represented in the factorized form as:
\begin{equation} \label{eq:1}
\mathbfcal{A}(l_1,l_2,\dots,l_d) \xlongequal{TTF}\mathcal{F}_{1}(l_1)\mathcal{F}_2(l_2)...\mathcal{F}_d(l_d)
\end{equation}
where $\mathcal{F}_k \in \mathbb{R}^{t_k\times r_{k-1}\times r_k}$, $l_k \in [1,t_k] \forall k\in [1,d]$. It implies that each entry in the tensor $\mathbfcal{A}$ is represented as a sequence of matrix multiplications. The set of tensors $\{\mathcal{F}_k\}_{k=1}^{d}$ are called core-tensors. The complexity of the TTF is determined by the sequence of ranks $\{r_0, r_1\dots,r_d\}$, where $r_0$ and $r_d$ must be set to  $1$ to obtain the outcome of the chain of multiplications in Eq. \ref{eq:1} always as a scalar.

If a contraint is imposed such a way that each integer $t_k$  as in Eq. \ref{eq:1} can be factorized as $t_k$ $=$ $p_k.q_k \forall k \in[1,d]$ and consequently each $\mathcal{G}_k$ is reshaped as  $\mathcal{G}^{*}_k$ $\in \mathbb{R}^{p_k\times q_k\times r_{k-1}\times r_k}$. Correspondingly, the factorization for the tensor $\mathbfcal{A}$ can be rewritten equivalently
to Eq.(1) as follows \cite{novikov2015tensorizing}:
\begin{equation} \label{eq:2}
\begin{split}
&\mathbfcal{A}((i_1,j_1),(i_2,j_2),\dots,(i_d,j_d)) \\ & \xlongequal{TTF}\mathcal{F}^{*}_1(i_1,j_1)\mathcal{F}^{*}_2(i_2,j_2)\dots\mathcal{F}^{*}_d(i_d,j_d)
\end{split}
\end{equation}
where the indices $i_k$ $=$ $\floor*{\frac{i_k}{q_k}} $, $j_k$ $=$ $l_k - q_k\floor*{\frac{i_k}{q_k}} $ and  $\mathcal{G}^{*}_k(i_k,j_k) \in \mathbb{R}^{r_{k-1}\times r_k}$. 
\subsection{Representation of feed-forward layer using TTF}
The fully connected layers of Neural network consist of several linear transformations as:
\begin{equation}
\mathbf{y}=\mathbf{W.x+b}
\end{equation} 
where $\mathbf{W}\in \mathbb{R}^{P\times Q}$, $\mathbf{x}\in \mathbb{R}^{P}$ and $\mathbf{y}\in \mathbb{R}^{Q}$. Eq.(3) can be written in scalar format as
\begin{equation}
\mathbf{y}(j)=\sum_{i=1}^{P}\mathbf{W}(i,j)\mathbf{x}(i)+\mathbf{b}(j) \hspace{2mm}\forall j \in [1,Q]
\end{equation}
Let us assume that $P$ and $Q$ are factorized into two integer arrays of the same length such as $P$=$\prod_{k=1}^{d}{p_k}$ and $Q$=$\prod_{k=1}^{d}{q_k}$. Based on this assumption, one can write the mapping function of tensors $\mathbfcal{X}$ $\in$ $\mathbb{R}^{p_1 \times p_2\dots\times p_d }\Rightarrow \mathbfcal{Y}$ $\in$ $\mathbb{R}^{q_1 \times q_2\dots\times q_d }$ as
\begin{equation}
\begin{split}
& \mathbfcal{Y}(j_1,j_2,\dots,j_d) \xlongequal{TTF} \\ & \sum_{i_1=1}^{m_1}\sum_{i_2=1}^{m_2}\dots\sum_{i_d=1}^{m_d}\mathbfcal{W}((i_1,j_1),(i_2,j_2),\dots,(i_d,j_d)).\\ &\mathbfcal{X}(i_1,i_2,\dots,i_d) + \mathbfcal{B}(j_1,j_2,\dots,j_d)
\end{split}
\end{equation}
The $d$-dimensional double-indexed tensor of weights $\mathbfcal{W}$ in Eq.(5) can be replaced by its TTF representation:
\begin{equation} \label{eq:6}
\begin{split}
&\mathbfcal{W}((i_1,j_1),(i_2,j_2),\dots,(i_d,j_d)) \\ & \xlongequal{TTF}\mathcal{F}^{*}_1(i_1,j_1)\mathcal{F}^{*}_2(i_2,j_2)\dots\mathcal{F}^{*}_d(i_d,j_d)
\end{split}
\end{equation}
Now instead of explicitly storing the full tensor $\mathbfcal{W}$ of size $\prod_{k=1}^{d}p_k.q_k$ $=$ $PQ$, only its TT format,i.e., the set of low-rank core tensors  $\{\mathcal{F}_k\}_{k=1}^{d}$ of size $\prod_{k=1}^{d}p_k.q_kr_{k-1}r_k$  can be stored. Hence, using TTF, the number of weight matrix parameters can be reduced and the compression rate can be given by:
\begin{equation}
C_r = \dfrac{\sum_{k=1}^{d}p_k.q_kr_{k-1}r_k}{\prod_{k=1}^{d}p_k.q_k}=\dfrac{\prod_{k=1}^{d}p_k.q_kr_{k-1}r_k}{PQ}
\end{equation}
When the weight matrix of a fully-connected layer is transformed into the tensor-train format, it produces a $Tensor$-$Train$ $Layer (TTL)$. We will use the following notation:
\begin{equation}
\mathbf{y}=TTL(\mathbf{W,x}) + \mathbf{b}
\end{equation} 
The TTL is compatible with the existing training
algorithms (such as stochastic gradient descent and its variant) for neural networks because all the derivatives required by the back-propagation algorithm  can be computed using the properties of the TT-format \cite{novikov2015tensorizing}.
\subsection{Tensor-train LSTM}
We incorporated TTF representation in a LSTM unit which can able to model long-term temporal dependencies even for the high-dimensional sequential data. As shown in Fig. 1, an LSTM unit contain a memory cell and three gates. The forget gate controls how much previous information should be erased from the cell and the input gate controls how much information should be added to the cell. Using TTL as in Eq.(8), the tensor-train factorized LSTM model can be described as follows:
\begin{equation}
\mathbf{i^{<t>}}=\sigma{(TTL(\mathbf{W}_i,[h^{<t-1>},x^{<t>}])+\mathbf{b}_i)}
\end{equation}
\begin{equation}
\mathbf{f^{<t>}}=\sigma{(TTL(\mathbf{W}_f,[h^{<t-1>},x^{<t>}])+\mathbf{b}_f)}
\end{equation}
\begin{equation}
\mathbf{o^{<t>}}=\sigma{(TTL(\mathbf{W}_o,[h^{<t-1>},x^{<t>}])+\mathbf{b}_o)}
\end{equation}
\begin{equation}
\tilde{\mathbf{c}}^{<t>}=tanh(TTL(\mathbf{W}_c,[h^{<t-1>},x^{<t>}])+\mathbf{b}_c)
\end{equation}
\begin{equation}
\mathbf{c}^{<t>}=\mathbf{i}^{<t>}*\tilde{\mathbf{c}}^{<t>}+\mathbf{f}^{<t>}*\mathbf{c}^{<t-1>}
\end{equation}
\begin{equation}
\mathbf{h}^{<t>}=\mathbf{o}^{<t>}*tanh(\mathbf{c}^{<t>})
\end{equation}
\begin{figure}[h]
  \centering
  \includegraphics[scale = 0.3]{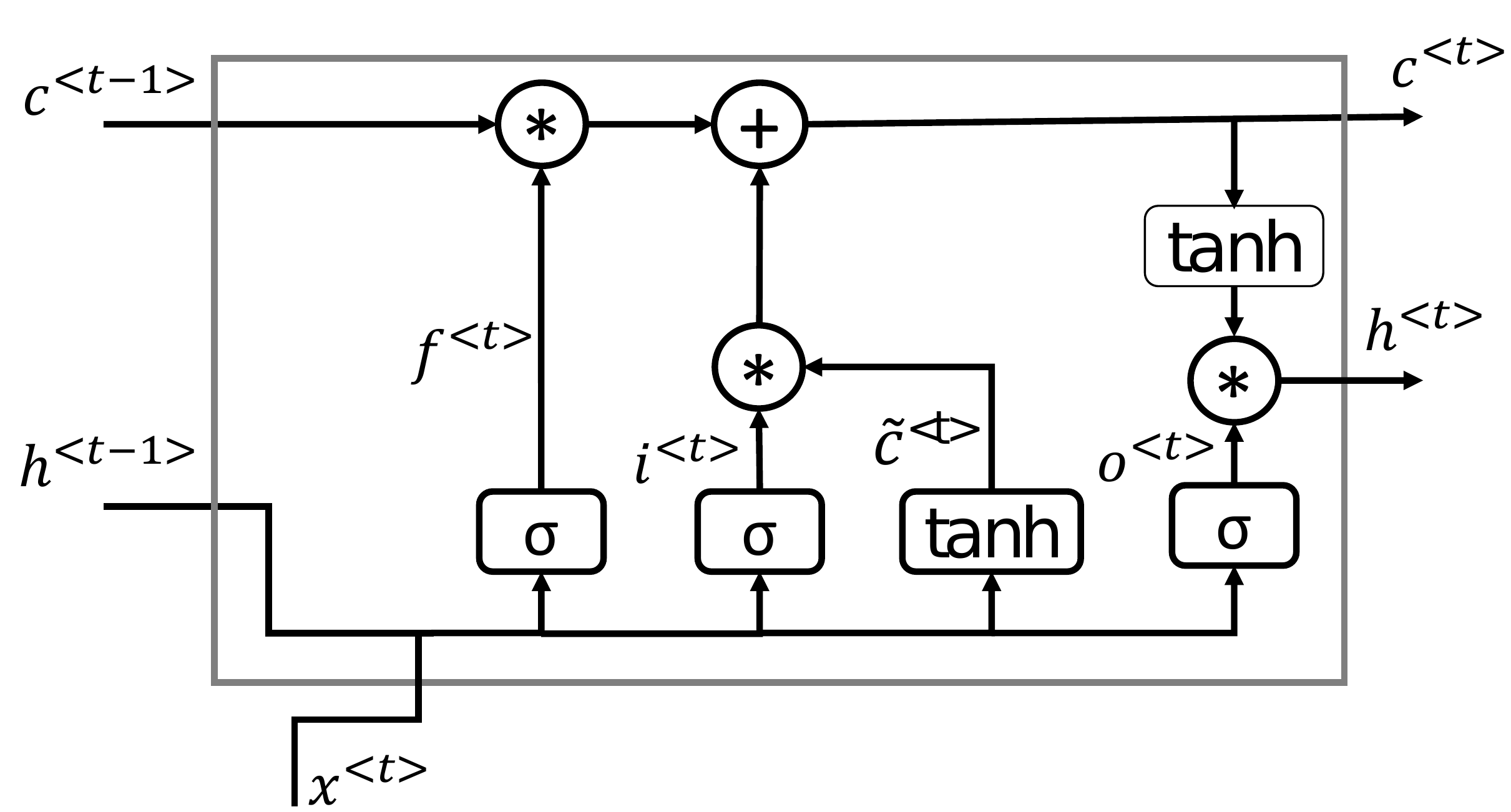}
  \caption{Block diagram of a basic LSTM unit.}
  \label{fig1}
\end{figure}
where $\mathbf{i^{<t>}}$, $\mathbf{f^{<t>}}$ and $\mathbf{o^{<t>}}$ denote input, forget and output gates respectively. $\mathbf{W}_i$, $\mathbf{W}_f$, $\mathbf{W}_o$ and $\mathbf{W}_c \in\mathbb{R}^{H\times (H+D)}$, represent the concatenated matrices which is required to compute each gate output. $H$ and $D$ denote the number of hidden units in LSTM block and the feature dimension of input $x^{<t>}$.   Hence, each LSTM block requires 4 TTL to represent its functionality in TTF format. It is possible to concatenate all the gates as one output tensor, which in turn parallelize the computation. It helps to  reduce further the number of parameters, where the concatenation is actually participating in the tensorization. The compression rate for the input-to-hidden weight matrix of LSTM now becomes  
\begin{equation}
C'_{r} = \dfrac{\sum_{k=1}^{d}p_kq_kr_{k-1}r_k + 3.p_1q_1r_0r_1}{4.\prod_{k=1}^{d}p_k.q_k}
\end{equation}
Bias vectors had not been converted into TTF representation because the number of bias parameters is insignificant compared to the number of parameters in matrices. In terms of performance,
the element-wise sum operation for bias vector is also insignificant compared to the matrix multiplication between a weight matrix and the input layer or the previous hidden layer.
\section{Speech enhancement framework}
A time-frequency (T-F) masking based supervised speech enhancement framework is adopted in the current work. We used a deep TT-LSTM based tensor-network which is trained in the supervised fashion to estimate a ideal-ratio mask (IRM) from the T-F feature representation of noisy speech. The audio mixtures  are first fed into a 64-channel gammatone filterbank, with center frequencies equally spaced in the range of  50 Hz to 8000 Hz on the $equivalent$ $rectangular$ $bandwidth$ (ERB) rate scale.  The output in each channel is then segmented into 20 ms frames with 50\% overlapping between successive frames. This way of processing would produce a time-frequency representation of the audio mixture, termed as  $Cochleagram$. 
The IRM can be defined as \cite{wang2014training}:
\begin{equation}
IRM(p,k) = \bigg {(}\frac{||S(p,k)||^2}{||S(p,k)||^2+ ||W(p,k)||^2}\bigg {)}^{\beta}
\end{equation} 
where $||S(p,k)||^2$ and $||W(p,k)||^2$  denote the clean speech energy and the noise energy, respectively, contained within T-F unit ($p$, $k$). $\beta$ is a tunable parameter and it has been empirically set to 0.5.

We have used Multi-Resolution Cochleagram (MRCG) \cite{zhang2014boosted} feature, extracted from the noisy speech, to train the model. 
In addition, velocity (delta feature) and acceleration (double-delta feature) coefficients are  appended with the 256-D raw features, resulting in 768-D feature vector from each frame.

Our deep tensor-net model has three TT-LSTM layers with 512 hidden units in each layer and  one fully-connected dense layer with 128 hidden units as shown in Fig. 3. We have used rectified linear units (ReLUs) for all intermediate hidden layers and dense layer. The output layer uses sigmoid activation since a ratio mask has the value in the range of [0,1]. 
\begin{figure}[h]
  \centering
  \includegraphics[scale = 0.22]{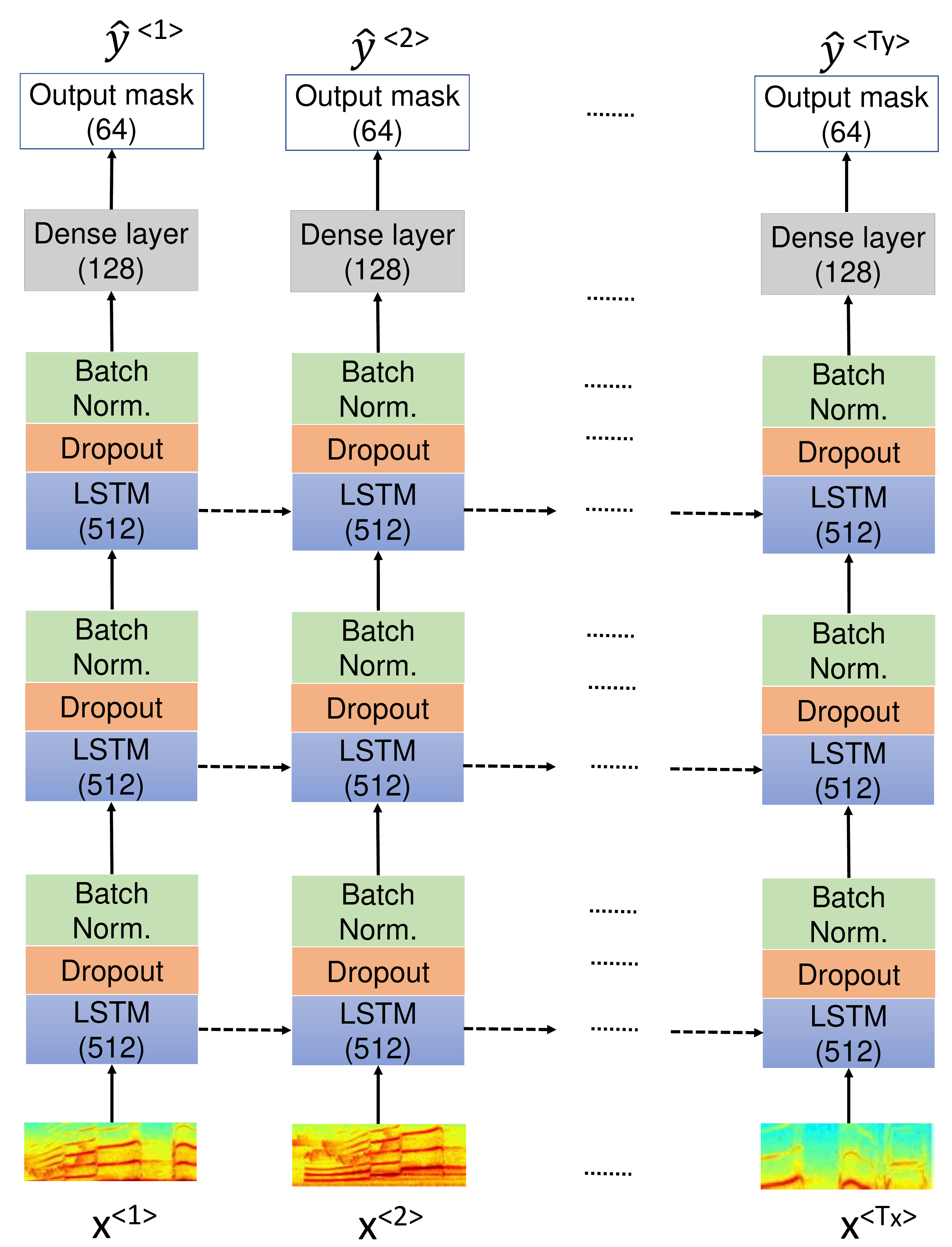}
  \caption{The TT-LSTM based tensor-net framework for speech enhancement.}
  \label{fig1}
\end{figure}
The dense and output layer is also represented in TT-format. For instance, the input feature dimension at each time step is 768 which is factorized as 16$\times$16$\times$3, the hidden layer is chosen to be 16$\times$16$\times$2 = 512 and the Tensor-Train ranks are [1, 4, 4, 4, 1], the first TT-LSTM layer in the framework will have only 10264 parameters including bias vectors, whereas A normal LSTM layer would have required 2,623,488 parameters to learn. The number parameters in each layer for the given framework in Figure 2 is summarized for both the normal LSTM based and TT-LSTM (with TT-Rank 4). The given tensor-net consists of three TT-LSTM layer with 512 (16$\times$16$\times$2) hidden units and TT-Rank 4 is denoted as TT-LSTM-H512-R-4 model.   We can also define TT-rank  and treat them as a hyper-parameter. In general,  the LSTM with a smaller TT-rank can yield a more efficient model but this would restrict the model to learn more complex representation. If we use a larger TT-rank, we get more flexibility to express our weight parameters but we sacrifice model efficiency at the same time.

\begin{table}[]
\centering
\scriptsize
\renewcommand{\arraystretch}{1.4}
\caption{Comparison of number of parameters and compression rate for normal neural network and tensor net with TT-rank=4 }

\label{my-label}
\begin{tabular}{c|c|c|c}
\hline \hline
\textbf{Layers}                                                     & \multicolumn{1}{l|}{\textbf{Normal  network}} & \multicolumn{1}{l|}{\textbf{Tensor net}} & \multicolumn{1}{l}{\textbf{Compression rate}} \\ \hline
\begin{tabular}[c]{@{}c@{}}\textbf{Layer 1} \\ (LSTM-RNN layer)\end{tabular} & 2,623,488                                     & 10,264                                   & $3.9x10^{-3}$                      \\ \hline
\begin{tabular}[c]{@{}c@{}}\textbf{Layer 2}\\ (LSTM-RNN layer)\end{tabular}  & 2,099,200                                     & 10,255                                   & $4.88x10^{-3}$                     \\ \hline
\begin{tabular}[c]{@{}c@{}}\textbf{Layer 3}\\ (LSTM-RNN layer)\end{tabular}  & 2,099,200                                     & 10,255                                   & $4.88x10^{-3}$ \\ \hline
\begin{tabular}[c]{@{}c@{}}\textbf{Layer 4}\\ (Dense layer)\end{tabular}     & 65,664                                        & 1,472                                    &  $2.2x10^{-2}$ \\ \hline
\begin{tabular}[c]{@{}c@{}}\textbf{Layer 5}\\ (Output layer)\end{tabular}    & 8,256                                         & 512                                      & $6.2x10^{-2}$                      \\ \hline
Total \# parameters                                                 & 6,895,808                                     & 32,760                                   & $4.75x10^{-3}$  \\ \hline \hline
\end{tabular}
\end{table}

\section{Experimental result}
To evaluate the performance of the proposed TT-LSTM based speech enhancement framework, a series of experiments have been conducted involving multiple matched and mismatched noise types at various SNR levels. 
\begin{table*}[]
\centering
\scriptsize
\renewcommand{\arraystretch}{1.2}
\caption{Evaluation  on Test Set A and Test Set B:  Objective measure represented as PESQ(STOI) score}
\label{my-label}
\begin{tabular}{||c|c|c|c|c|c||}
\hline \hline
\textbf{\begin{tabular}[c]{@{}c@{}}SNR\\ (dB)\end{tabular}} & \textbf{\begin{tabular}[c]{@{}c@{}}Noisy\\ (Unprocessed)\end{tabular}} & \textbf{\begin{tabular}[c]{@{}c@{}}Proposed \\ TT-LSTM-Rank-4\end{tabular}} & \textbf{\begin{tabular}[c]{@{}c@{}}Deep LSTM [7] \end{tabular}} & \textbf{\begin{tabular}[c]{@{}c@{}}DNN-based \\ Regression model [14] \end{tabular}} & \textbf{\begin{tabular}[c]{@{}c@{}}DNN-based \\ T-F masking [12] \end{tabular}} \\ \hline
-6                                                          & 1.42(0.702)                                                       & \textbf{1.98(0.762)}                                                                          & 1.89(0.772)                                                                 & 1.64(0.756)       & 1.52(0.712)   \\ \hline
-3                                                          & 1.78(0.794)                                                       & \textbf{2.46(0.849)}                                                                          & 2.56(0.859)                                                                 & 2.06(0.792)       & 2.36(0.707)   \\ \hline
0                                                           & 2.06(0.762)                                                       & \textbf{2.78(0.816)}                                                                          & 2.88(0.825)                                                                 & 2.28(0.712)       & 2.46(0.752)   \\ \hline
3                                                           & 2.56(0.812)                                                       & \textbf{3.59(0.822)}                                                                          & 3.82(0.832)                                                                 & 2.66(0.749)       & 2.93(0.729)   \\ \hline
6                                                           & 3.16(0.842)                                                       & \textbf{3.62(0.882)}                                                                          & 3.76(0.898)                                                                 & 2.76(0.782)       & 3.14(0.712)   \\ \hline
9                                                           & 3.47(0.872)                                                       & \textbf{3.96(0.927)}                                                                          & 4.12(0.892)                                                                 & 3.26(0.832)       & 3.21(0.818)   \\ \hline \hline
\end{tabular}
\end{table*}
\subsection{Audio material: speech and noise corpus}
The clean speech material are taken from the TIMIT \cite{garofolo1993darpa} and Voice Bank corpus \cite{veaux2013voice}. The clean sentences used for training and validation are taken by randomly selecting utterances of 20 speakers (10 male and 10 female) of Voice Bank database and 90 speakers (45 male and 45 female) from TIMIT corpus. The training and validation sets consist of 5250 and 2750 utterances, respectively, which is equivalent to approximately 10 hours of training data and 2.5 hours of validation data. To create noisy stimuli for training, we used eight different types of noise instances: synthetically generated noises (speech-shaped noise (SSN) and six-speaker babble noise) and six real-world recordings from the DEMAND database \cite{thiemann2013diverse}. The noise chosen from DEMAND database are: domestic noise (inside a kitchen), an office noise (inside a meeting room), two public space noises ( cafeteria and subway station) and two transportation noises (car and metro). All the utterances and noise samples are down-sampled to 16 kHz. Each clean sentence is embedded in the aforementioned background noises at SNRs ranging from -6 dB to +9 dB (with 3 dB step). To achieve a desired SNR, the noise signal is scaled based on the active speech level of the clean speech signal as per ITU P.56 \cite{recommendation1993objective}.

Our test set contains 300 utterances of 12 speakers (6 male and 6 female) from Voice Bank corpus and 1120 utterances of 56 speakers (28 male and 28 female) from TIMIT database. As we have given emphasis on the speaker generalization in the evaluation process, the speakers involved in the test set are entirely different from the speakers in the train and validation (dev) set. To conduct assessment in both matched and mismatched noise scenarios, the following two separate test sets are generated:
\begin{itemize}
\item \textbf{Test set A}: Each sentence of test set is mixed with four types of noise (SSN, cafeteria, car, subway station) which are already seen during training time.
\item \textbf{Test set B}: Each sentence of test set is mixed with two types of noise (washroom and restaurent noise) taken from DEMAND corpus and other two types of noise (pink and factory) from NOISEX-92 corpus \cite{varga1993assessment}. This is a test set containing mismatched noise which are completely unseen during training time.
\end{itemize} 
\subsection{Experimental setup and model training}
For evaluating the performance of the proposed speech enhancement framework using TT-LSTM based TensorNet, we have trained different models with the same architecture as shown in Figure 3, but having different configurations of TT-LSTM, particularly varying the TT-Ranks. Here, we have presented the result considering a system with TTF-Rank = 4, denoted by TT-LSTM-Rank-4. 
We have also considered the baseline DNN and normal LSTM based speech enhancement framework with 3 layers of hidden layers and having 512 hidden units in each layer. For training, the back-propagation through time has been used. Dropout (probability = 0.5) and batch-normalization also has been employed as it is effective to achieve better generalization.  We compared the performance of these systems on the two test sets. The perceptual evaluation of speech quality (PESQ) (wide-band) \cite{recommendation2001perceptual}\cite{rec2005p} is used to evaluate the proposed systems in terms of perceived speech quality improvement while the short-time objective intelligibility (STOI) \cite{taal2011algorithm} measure is adopted to predict the speech intelligibility. 
The comparative performance of different models in terms of PESQ and STOI are illustrated in Table II. It is evident form the average result on Test set-A and Test set-B, TT-LSTM based Tensornet performs almost similar even though its complexity in terms of number of parameters is orders of magnitude less.  
It indicates that TT-LSTM and LSTM both perform significantly better than the other DNN based speech enhancement framework. Moreover, TT-LSTM can achieve competitive performances with the state-of-the-art uncompressed RNN model, even though the proposed model architecture is orders of magnitude less complex.
%
%
%
%
%
%
%
%
\section{Conclusions}
In this letter, a tensor-train factorized LSTM model has been introduced in the context of supervised single-channel speech enhancement application. Tensor-Train (TT) format is an effective way of reducing parameters by representing the weight matrices of LSTM. Based on this TT-LSTM units, we proposed a T-F masking based deep TensorNet model for single-channel speech enhancement task. Experimental results in various test conditions and in terms of standard speech quality and intelligibility metrics, demonstrated that the proposed deep TT-LSTM based speech enhancement framework can achieve competitive performances with the state-of-the-art uncompressed RNN model, even though the proposed model architecture is orders of magnitude less complex. 


\bibliographystyle{IEEEtran}

\bibliography{ref_tt_lstm}

\begin{thebibliography}{10}
\providecommand{\url}[1]{#1}
\csname url@samestyle\endcsname
\providecommand{\newblock}{\relax}
\providecommand{\bibinfo}[2]{#2}
\providecommand{\BIBentrySTDinterwordspacing}{\spaceskip=0pt\relax}
\providecommand{\BIBentryALTinterwordstretchfactor}{4}
\providecommand{\BIBentryALTinterwordspacing}{\spaceskip=\fontdimen2\font plus
\BIBentryALTinterwordstretchfactor\fontdimen3\font minus
  \fontdimen4\font\relax}
\providecommand{\BIBforeignlanguage}[2]{{%
\expandafter\ifx\csname l@#1\endcsname\relax
\typeout{** WARNING: IEEEtran.bst: No hyphenation pattern has been}%
\typeout{** loaded for the language `#1'. Using the pattern for}%
\typeout{** the default language instead.}%
\else
\language=\csname l@#1\endcsname
\fi
#2}}
\providecommand{\BIBdecl}{\relax}
\BIBdecl

\bibitem{loizou2013}
P.~C. Loizou, \emph{Speech enhancement: theory and practice}.\hskip 1em plus
  0.5em minus 0.4em\relax CRC press, 2013.

\bibitem{sie2017}
M.~Kolbæk, Z.~H. Tan, and J.~Jensen, ``Speech intelligibility potential of
  general and specialized deep neural network based speech enhancement
  systems,'' \emph{IEEE/ACM Transactions on Audio, Speech, and Language
  Processing}, vol.~25, no.~1, pp. 153--167, Jan 2017.

\bibitem{chen2017long}
J.~Chen and D.~Wang, ``Long short-term memory for speaker generalization in
  supervised speech separation,'' \emph{The Journal of the Acoustical Society
  of America}, vol. 141, no.~6, pp. 4705--4714, 2017.

\bibitem{hochreiter1997long}
S.~Hochreiter and J.~Schmidhuber, ``Long short-term memory,'' \emph{Neural
  computation}, vol.~9, no.~8, pp. 1735--1780, 1997.

\bibitem{greff2017lstm}
K.~Greff, R.~K. Srivastava, J.~Koutn{\'\i}k, B.~R. Steunebrink, and
  J.~Schmidhuber, ``Lstm: A search space odyssey,'' \emph{IEEE transactions on
  neural networks and learning systems}, vol.~28, no.~10, pp. 2222--2232, 2017.

\bibitem{graves2013speech}
A.~Graves, A.-r. Mohamed, and G.~Hinton, ``Speech recognition with deep
  recurrent neural networks,'' in \emph{Acoustics, speech and signal processing
  (icassp), 2013 ieee international conference on}.\hskip 1em plus 0.5em minus
  0.4em\relax IEEE, 2013, pp. 6645--6649.

\bibitem{kalchbrenner2013recurrent}
N.~Kalchbrenner and P.~Blunsom, ``Recurrent continuous translation models,'' in
  \emph{Proceedings of the 2013 Conference on Empirical Methods in Natural
  Language Processing}, 2013, pp. 1700--1709.

\bibitem{lipton2015critical}
Z.~C. Lipton, J.~Berkowitz, and C.~Elkan, ``A critical review of recurrent
  neural networks for sequence learning,'' \emph{arXiv preprint
  arXiv:1506.00019}, 2015.

\bibitem{samui2017deep}
S.~Samui, I.~Chakrabarti, and S.~K. Ghosh, ``Deep recurrent neural network
  based monaural speech separation using recurrent temporal restricted
  boltzmann machines,'' \emph{Proc. Interspeech 2017}, pp. 3622--3626, 2017.

\bibitem{samui2018asoc}
------, ``Time-frequency masking based supervised speech enhancement framework
  using fuzzy deep belief network,'' \emph{Applied Soft Computing}, 2018.

\bibitem{oseledets2011tensor}
I.~V. Oseledets, ``Tensor-train decomposition,'' \emph{SIAM Journal on
  Scientific Computing}, vol.~33, no.~5, pp. 2295--2317, 2011.

\bibitem{novikov2015tensorizing}
A.~Novikov, D.~Podoprikhin, A.~Osokin, and D.~P. Vetrov, ``Tensorizing neural
  networks,'' in \emph{Advances in Neural Information Processing Systems},
  2015, pp. 442--450.

\bibitem{wang2014training}
Y.~Wang, A.~Narayanan, and D.~Wang, ``On training targets for supervised speech
  separation,'' \emph{Audio, Speech, and Language Processing, IEEE/ACM
  Transactions on}, vol.~22, no.~12, pp. 1849--1858, 2014.

\bibitem{zhang2014boosted}
X.-L. Zhang and D.~Wang, ``Boosted deep neural networks and multi-resolution
  cochleagram features for voice activity detection,'' in \emph{Fifteenth
  Annual Conference of the International Speech Communication Association},
  2014.

\bibitem{garofolo1993darpa}
J.~S. Garofolo, L.~F. Lamel, W.~M. Fisher, J.~G. Fiscus, and D.~S. Pallett,
  ``Darpa timit acoustic-phonetic continous speech corpus cd-rom. nist speech
  disc 1-1.1,'' \emph{NASA STI/Recon technical report n}, vol.~93, 1993.

\bibitem{veaux2013voice}
C.~Veaux, J.~Yamagishi, and S.~King, ``The voice bank corpus: Design,
  collection and data analysis of a large regional accent speech database,'' in
  \emph{Oriental COCOSDA held jointly with 2013 Conference on Asian Spoken
  Language Research and Evaluation (O-COCOSDA/CASLRE), 2013 International
  Conference}.\hskip 1em plus 0.5em minus 0.4em\relax IEEE, 2013, pp. 1--4.

\bibitem{thiemann2013diverse}
J.~Thiemann, N.~Ito, and E.~Vincent, ``The diverse environments multi-channel
  acoustic noise database: A database of multichannel environmental noise
  recordings,'' \emph{The Journal of the Acoustical Society of America}, vol.
  133, no.~5, pp. 3591--3591, 2013.

\bibitem{recommendation1993objective}
I.~Recommendation, ``Objective measurement of active speech level,''
  \emph{ITU-T Recommendation}, p.~56, 1993.

\bibitem{varga1993assessment}
A.~Varga and H.~J. Steeneken, ``Assessment for automatic speech recognition:
  Ii. noisex-92: A database and an experiment to study the effect of additive
  noise on speech recognition systems,'' \emph{Speech communication}, vol.~12,
  no.~3, pp. 247--251, 1993.

\bibitem{recommendation2001perceptual}
I.-T. Recommendation, ``Perceptual evaluation of speech quality (pesq): An
  objective method for end-to-end speech quality assessment of narrow-band
  telephone networks and speech codecs,'' \emph{Rec. ITU-T P. 862}, 2001.

\bibitem{rec2005p}
I.~Rec, ``P. 862.2: Wideband extension to recommendation p. 862 for the
  assessment of wideband telephone networks and speech codecs,''
  \emph{International Telecommunication Union, CH--Geneva}, 2005.

\bibitem{taal2011algorithm}
C.~H. Taal, R.~C. Hendriks, R.~Heusdens, and J.~Jensen, ``An algorithm for
  intelligibility prediction of time--frequency weighted noisy speech,''
  \emph{Audio, Speech, and Language Processing, IEEE Transactions on}, vol.~19,
  no.~7, pp. 2125--2136, 2011.

\end{thebibliography}

\end{document}